# A practical omni-directional SH wave transducer for structural health monitoring based on two thickness-poled piezoelectric half-rings

Qiang Huan[1,2], Mingtong Chen[1], Faxin Li[1,2,3,a]

[1] LTCS and College of Engineering, Peking University, Beijing 100871, China
[2] Center for Applied Physics and Technology, Peking University, Beijing, China
[3] Beijing Key Laboratory of Magnetoelectric Materials and Devices, Peking University, Beijing, China

**Abstract:** Structural health monitoring (SHM) has become more and more important in modern industries as it can monitor the safety of structures during the full service life and prevent possible losses of life and economics. Shear horizontal wave in plate-like structures is very useful for long distance defects inspection since its fundamental mode ($SH_0$) is totally non-dispersive. However, all the currently available SH wave transducers are not suitable for practical SHM. In this work, we firstly investigated via finite element simulations the performances of thickness-poled $d_{15}$ PZT ring based omni-directional SH wave piezoelectric transducers (OSH-PT) consisting of different number of elements. Results show that the two half-rings based OSH-PT can have perfect omni-directivity and acceptable performances in excitation/reception of $SH_0$ waves, and its performances can be fairly enhanced by reducing the outer and inner diameters. Experimental results on a 2mm-thick, 21mm outer-diameter, 9mm inner-diameter two-half-ring OSH-PT shows that it has good omni-directional properties with the maximum deviation of ~14% in both excitation and reception of $SH_0$ wave. The signal to noise ratio (SNR) of the OSH-PT in the case of self-excitation and self-reception is over 15dB from 145kHz to 200kHz, which is acceptable for most applications. The proposed two-half-ring OSH-PT is expected to be widely used in $SH_0$ wave based SHM due to its simple structure, easy fabrication/assembling, low cost and acceptable performances.

---

[a] Author to whom all correspondence should be addressed, Email: lifaxin@pku.edu.cn

# 1. Introduction

Nondestructive testing (NDT) and structural health monitoring (SHM) have been widely used in modern industries as they can evaluate the safety of structures/systems in a nondestructive manner and prevent the possible losses of life and economics[1,2]. The main difference between NDT and SHM is that whether the sensors/probes are movable or fixed on the structures. With the rapid development of wireless sensor technology and computational capacity of personal computers, the applications of SHM turns to be realistic and have been paid more and more attention to in recent years. If it is feasible, SHM is obviously superior to NDT since it can monitor the conditions of structures for the whole service life and remote monitoring is also possible[3]. In a SHM system, sensors or transducers are the most important component, which should be reliable, cost effective, of compact size and easy to assemble.

In the past two decades, guided wave based techniques have been widely used and intensively studied in SHM of large structures[4–8]. Guided waves, e.g., Lamb waves and shear horizontal (SH) waves in plate-like structures, have the advantage of long-distance propagation with small attenuation. So far, Lamb wave based SHM technique have been well investigated in isotropic thin plates and is still under development for layered composites[7]. The rapid development of Lamb wave based SHM is due to the nearly perfect properties of PZT ceramics. A circular thickness-poled PZT wafer is inherently an omni-directional Lamb wave transducer, which is of high signal to noise ratio (SNR), cost effective, small and easy to be bonded on plates. However, Lamb waves are inherently dispersive, of multi-mode and encounter mode conversions at defects or boundaries, introducing difficulties in explanation of the received signals. Giurgiutiu proposed that quasi-single mode Lamb wave can be excited by frequency tuning[9]. However, at the tuned frequency, the quasi-single mode Lamb wave may still be dispersive. Furthermore, in practical SHM applications, many PZT wafers are required and it is not possible to ensure that all the wafers have the best tuning properties at a single frequency. Therefore, in Lamb wave based SHM systems, it is rather difficult to achieve a high SNR in defect imaging[10,11].

In comparison, the fundamental SH wave ($SH_0$) is totally non-dispersive and less mode conversion will occur when it encounters defects or boundaries. However, it is not straightforward to excite $SH_0$ wave in plates. Traditionally, $SH_0$ waves can be excited by using electromagnetic acoustic transducers (EMAT) in a non-contact manner which was firstly introduced by Thompson in late 1970s[12,13]. Later, the magnetostrictive patch transducers (MPT) were developed and had been used in pipe inspection since the coupling efficiency of MPT is much higher than that of EMAT[14,–15], and the fundamental torsional wave T(0,1) in pipes is analogous to $SH_0$ in plates, i.e., both are non-dispersive[16]. T(0,1) wave can also be excited/received by a thickness-shear ($d_{15}$) PZT based piezoelectric ring in a dry-coupled manner or bonded on the pipe[17–19]. However, a $d_{15}$ PZT wafer not only generates $SH_0$ waves, but also generate Lamb waves in plates simultaneously[20,21]. Recent investigations indicate that $SH_0$ wave can also be excited by using $d_{36}$ face-shear piezoelectric crystals and ceramics[22,23]. However, single mode $SH_0$ wave still cannot be excited until the recent narrow-band $SH_0$ wave excitation by using a synthetic $d_{36}$ mode PZT wafer[24]. Miao et al proposed another face-shear $d_{24}$ mode in PZT wafers and successfully excited single-mode $SH_0$ wave in a wide frequency range[25]. Furthermore, the $d_{24}$ mode PZT wafer can filter Lamb waves and receive $SH_0$ wave only in a wide frequency range. The $d_{24}$ mode PZT wafers was also assembled into a ring to excite and receive T(0,1) wave in pipes[26].

It should be noted that all the above-mentioned SH wave transducer are directional and cannot be directly used for SHM applications where omni-directional transducers are required. Omni-directional SH wave transducers were firstly realized by Kim et al using MPT[27] and later using EMAT[28]. Recently, Liu et al employed the MPT based omni-directional $SH_0$ wave transducer array to detect defects in quasi-isotropic composites[29]. However, due to the complicated structures and high power driving problem, both the omni-directional MPT and EMAT are not suitable for SHM applications. Borigo et al proposed a design of omni-directional SH wave piezoelectric transducer (OSH-PT) based on two circumferentially poled PZT half-ring[30]. However, although some PZT manufacturers claimed so, uniform circumferential polarization cannot be possibly realized in a half-ring because the induced polarization during poling is not proportional to the applied field. In practice, quasi-uniform circumferential polarization is usually synthesized by assembling several equal-sized in-plane poled PZT elements circumferentially, such as the Langevin torsional transducers[31]. Belanger and Boivin fabricated an OSH-PT by synthesizing circumferential polarization using six $d_{15}$ mode PZT wafers[32]. The sensitivity deviation in omni-directionally generating $SH_0$ waves is around 20% while the reception properties were not examined. Miao et al built an OSH-PT based on synthesized circumferential polarization using twelve $d_{24}$ mode PZT wafers[33]. The omni-directional sensitivity deviations in $SH_0$ wave generation and reception are both around 15%. It should be noted that uniform circumferential polarization is rather difficult to be synthesized because it is almost impossible to ensure that the polarization in each PZT wafer is exactly the same, although the size can be strictly controlled. Recently, Huan et al proposed an OSH-PT based on a thickness-poled, thickness-shear ($d_{15}$) PZT ring in which the electric field is applied circumferentially by evenly dividing the ring into twelve wafers[34]. Thanks to the uniform thickness-polarization in each wafers, the omni-directional sensitivity deviations in generating and receiving $SH_0$ waves are only 6-7%. The thickness-poled $d_{15}$ PZT ring based OSH-PTs have shown superior performances in defect imaging and localization, compared to the Lamb wave counterparts[35]. However, this OSH-PT still consists of twelve PZT elements and is not easy to fabricate/assemble, which is not suitable for practical SHM applications where usually a large number of transducers are required to be fixed on the structures.

In this work, we aimed to develop an OSH-PT for practical SHM applications which should be easily fabricated and assembled on structures. The primary candidate is the thickness-poled $d_{15}$ PZT ring equally dividing into several elements. Firstly, finite element (FEM) simulations were conducted to examine the performances of the thickness-poled $d_{15}$ PZT ring with different number of elements in excitation and reception of $SH_0$ waves. Results show that an OSH-PT consisting of two thickness-poled PZT half-rings can have uniform sensitivity along all directions. The simulated signal to noise ratio (SNR) in $SH_0$ wave excitation can reach 16 dB for large rings (outer-diameter, OD>21mm) and it steadily increases with the decreasing OD. Then, an OSH-PT based on two half-rings with the OD of 21mm was fabricated and its performances were systematically tested. Results show that this OSH-PT can excite and receive $SH_0$ wave with the SNR better than 15dB from 145kHz to 200kHz. The omni-directivity of the fabricated OSH-PT is also examined, with the maximum deviation of 14% in both excitation and reception, which is mainly due to the measurement limitations. The proposed two-half-ring OSH-PT is very easy to fabricate and assemble, which is expected to pave the road to practical SHM systems based on SH waves.

## 2. Methodology
### 2.1 Design of the omni-directional SH wave piezoelectric transducer (OSH-PT)

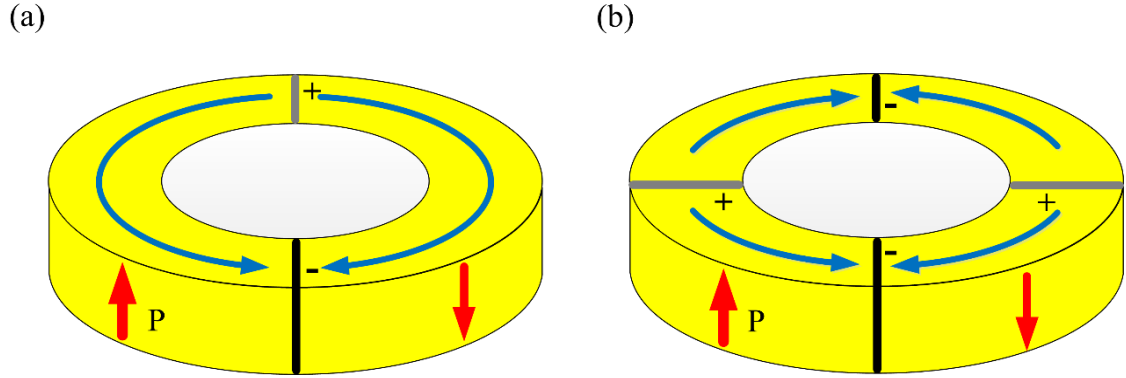

Fig.1. Design of practical OSH-PT based on thickness-poled $d_{15}$ mode PZT rings with different number of elements: (a) two half-rings; (b) four quarter-rings

The primary candidate for the practical OSH-PT is the thickness-poled $d_{15}$ mode PZT ring since such a ring consisting of twelve elements had been verified to have very good performances in excitation/reception of $SH_0$ wave omni-directionally[34]. Since a ring with twelve elements is not easy to fabricate or assemble, we want to examine the performances of such a ring with fewer elements in $SH_0$ wave excitation/reception. Obviously, a ring consisting of fewer elements is easier to fabricate and assemble. So if its performance is acceptable, it can turn to be a practical OSH-PT for SHM applications. Fig.1 shows the design of an OSH-PT based on thickness-poled $d_{15}$ mode PZT rings consisting of two half-rings and four quarter-rings. The PZT ring is firstly poled along the thickness direction which is easy to realize. After poling, the ring is cut into even-number elements along its diameter. Then the top/bottom electrodes were removed and lateral electrodes were spread for circumferential field loading. In assembling the OSH-PT, the poling directions of the adjacent elements were opposite thus they can share the common lateral electrodes.

**2.2 Finite Element Simulations**

Firstly, finite element (FEM) simulations were carried out in ANSYS to check the performances of OSH-PT based on thickness-poled $d_{15}$ mode PZT rings with different number of elements in excitation and reception of $SH_0$ waves. An aluminum plate (with the Young's modulus of 69 GPa, Poisson ratio of 0.33, and density of 2700 kg·m$^{-3}$) was modeled by element SOLID 185 with the dimensions of 400 mm × 400 mm × 2 mm. The OSH-PT modeled by elements SOLID 5 was bonded on the central of the plate, and its material was PZT-5H whose properties can be found elsewhere [36]. The dimensions of the ring are firstly fixed to be the same as that in previous studies[34] (with the outer diameter of 21mm, inner diameter of 9mm and thickness of 2mm) and is then varied for the two half-ring case to study the size effect. During the simulations, the excitation signal was modulated into a five cycle sinusoid tone burst enclosed in Hanning window with central frequency of 180 kHz. The drive voltage was set to be inversely proportional to the number of elements thus to keep the same electric field intensity for all the cases. For the two half-ring case, the drive voltage is 120V.

## 3. FEM Simulation Results and Discussion

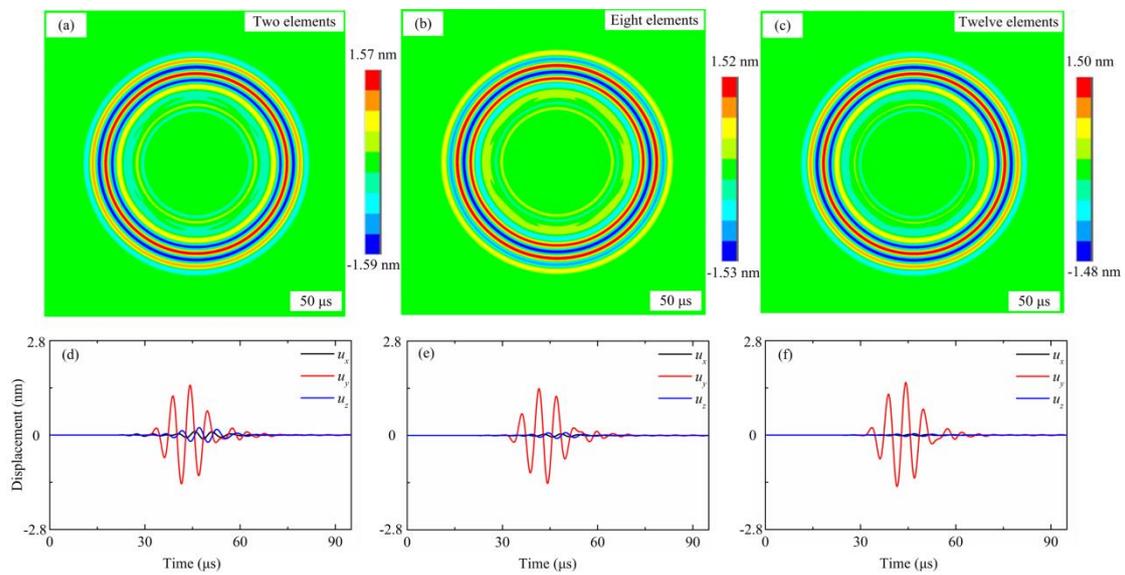

Fig.2. Comparisons of the OSH-PTs consisting of different number of elements in $SH_0$ wave excitation. Up): transient tangential displacement component at 50 μs; Bottom): Time-domain displacement components under cylindrical coordinates. Left): Two elements; Middle): eight elements; Right): Twelve elements.

The performances of the thickness poled $d_{15}$ mode PZT rings consisting of 2, 4, 6, 8, 10, 12 and 16 elements in excitation of $SH_0$ wave is firstly simulated. Due to the limited space, only the results for the two (half-rings), eight and twelve elements cases were plotted in Fig.2 Firstly, the tangential displacement dominated by $SH_0$ wave were compared for these three cases. The snapshots of the displacement fields at 50 μs were shown in Fig.2(a), (b), and (c). It can be found that all the three tangential displacement fields were perfectly axisymmetric, which means that $SH_0$ wave were generated omni-directionally with uniform amplitudes. The amplitude of $SH_0$ wave for these three cases is very close to each other. It slightly decreases with the increasing number of elements, which should be due to the reduced in-plane stiffness of the whole ring. When comparing the total displacement components in time-domain, the results were quite different, as seen in the bottom row of Fig.2 Besides the $SH_0$ wave ($u_y$), the $S_0$ wave (radial displacement component $u_x$) and $A_0$ wave (out-of-plane displacement component $u_z$) also appeared and $A_0$ is always dominant to $S_0$ for all these three cases. Meanwhile, it can be seen that the amplitudes of Lamb waves decrease with the increasing number of elements. This indicates that Lamb waves ($A_0$ wave in this work) can be better suppressed in an OSH-PT consisting of more elements where the out-of-plane stiffness of each element is fairly considerably enhanced.

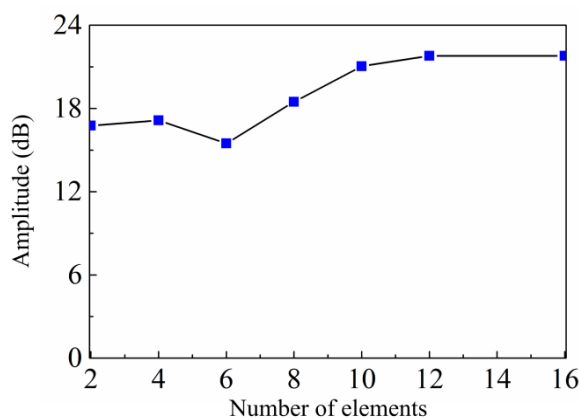

Fig.3. The simulated SNR of the excited $SH_0$ wave to Lamb waves by the OSH-PTs consisting of different number of elements

Fig.3 presented the simulated SNRs of the excited $SH_0$ wave to Lamb waves by the OSH-PT consisting of different number of elements. It can be seen that overall the SNR increases with the increasing number of elements, it is about 16.5dB for the two-half-ring case and saturates at about 22dB when the element number is over twelve. However, there is an exception for the six-element case whose SNR in excitation is only about 15dB, even lower than that for the two and four-elements case. The mechanism for this exception cannot be well clarified at present, while this may explain the reported poor SNR in excitation of $SH_0$ wave by an OSH-PT using six $d_{15}$ PZT elements to synthesize the circumferential polarization[32].

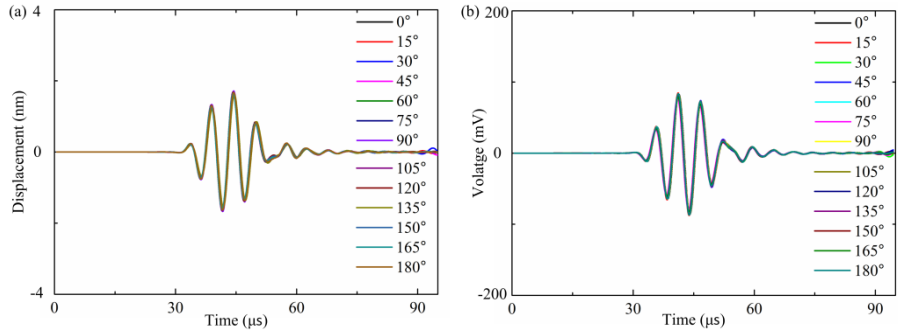

Fig.4. Time-domain FEM simulation results of the two-half-ring OSH-PT in exciting and receiving SH0 wave. (a) Tangential displacement excited by the OSH-PT at different directions; (b) voltage signals received by the OSH-PT at different directions.

Since the simulated SNR in $SH_0$ wave excitation by the two-element (two-half-ring) OSH-PT is very close to that by the four-element case, the two-half-ring OSH-PT is thus further investigated since it is the most convenient for fabrication and assembling. To further examine the omni-directional excitation/reception performances of the two-half-ring OSH-PT, time-domain FEM simulations was carried out in ANSYS. Centering at the OSH-PT, thirteen monitoring points were positioned at a 100mm-radium circle with the angle interval of 15°. When checking its performance in $SH_0$ wave generation, tangential displacements were picked up at the monitoring points. The results were shown in Fig.4(a). It can be seen that all signals along different directions were entirely overlapped, which further conformed the uniform omni-directivity of the OSH-PT in $SH_0$ wave generation. Meanwhile, the waveform of the received signals was the same as that of the input signal with little distortion. When checking the two-half-ring OSH-PT's performance in $SH_0$ wave reception, tangential displacements were applied at the monitoring points respectively to simulate the $SH_0$ wave and the OSH-PT served as a sensor. The received voltages were shown in Fig.4(b). Again, all signals were totally overlapped without waveform distortion. These results validated that theoretically the two-half-ring OSH-PT can generate and receive $SH_0$ wave omni-directionally with uniform sensitivity.

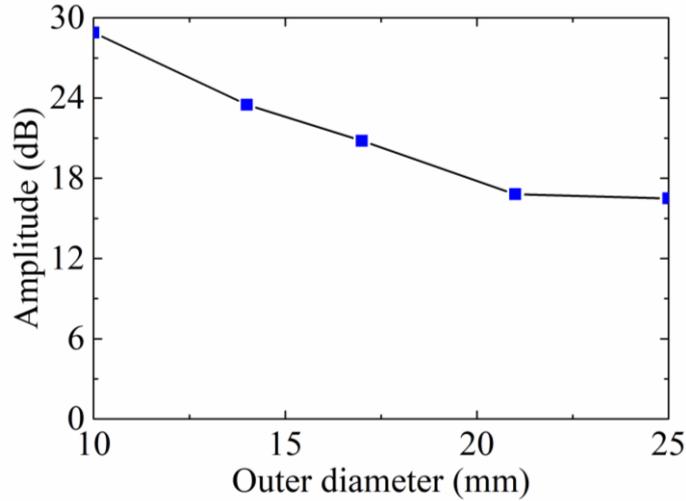

Fig.5 The simulated SNR of $SH_0$ wave to Lamb waves in excitation by 2mm-thick two-half-ring OSH-PTs with different outer diameters. The ratio of inner to outer diameter is fixed at 9/21.

We also conducted FEM simulations on the two-half-ring OSH-PTs with different outer diameters in $SH_0$ wave excitation. In the simulations, the ratio of inner diameter to outer diameter is fixed at 9/21, thus the above simulations results can be included for reference. The simulated SNR of $SH_0$ wave to Lamb waves is plotted in Fig.5. It can be seen that the SNR of the two-half-ring OSH-PT in $SH_0$ wave excitation increases steadily with the decreasing diameter, it can even reach 29dB when the outer diameter is reduced to 10mm.

## 4. Experimental

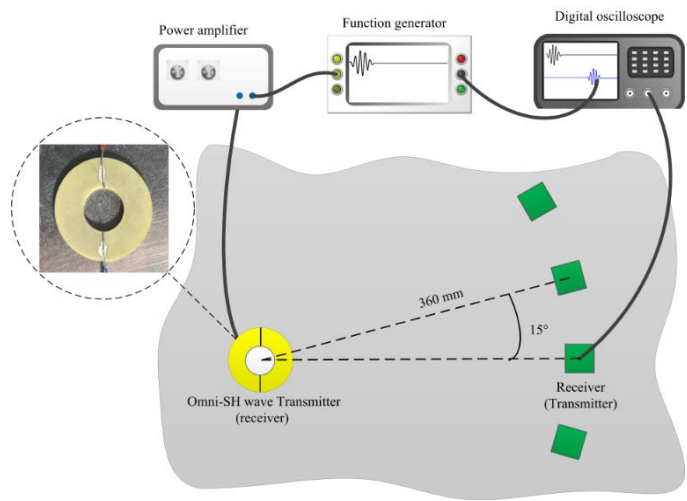

Fig.6. The schematic of testing setup to check the excited/received wave mode and omni-directivity of the two-half-ring OSH-PT.

Since the FEM simulation results show that the OSH-PT consisting of two half-rings can generate $SH_0$ wave with the SNR better than 16dB and the omni-directivity is theoretically perfect, we then fabricated the two-half-ring OSH-PT and examined its performances experimentally. The size of the tested OSH-PT is the same as what reported previously for comparison[35] (outer diameter of 21mm, inner diameter of 9mm and thickness of 2mm). The layout of the testing setup for $SH_0$ wave excitation and reception was shown in Fig.6. During the testing, the two-half-ring OSH-PT and several $d_{36}$ type PMN-PT wafers (which can excite/receive both $SH_0$ waves and Lamb waves) were bonded on a 1000mm × 1000mm × 2mm aluminum plate as actuator/sensors with the distances of 360mm. A five-cycle Hanning window-modulated sinusoid signal generated by a functional generator (Agilent 33220A) and amplified by a power amplifier (KH7602M) was used to drive the actuator. A digital oscilloscope (Agilent DSO-X 3024A) was used to collect the wave signals received by the sensors. In all the testing, the drive voltage for the two-half-ring OSH-PT is 120V and that for the $d_{36}$ type PMN-PT wafer is 50V.

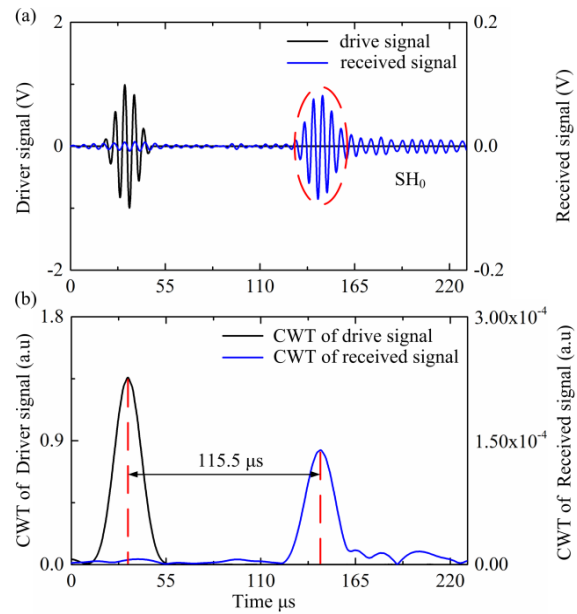

Fig.7. (a) Signals excited by the two-half-ring OSH-PT at 180 kHz with drive voltage of 120V and received by a $d_{36}$ type PMN-PT single crystal wafer. (b) Continuous wavelet transform (CWT) of the drive signal and received signal.

Firstly, the performance of the two-half-ring OSH-PT in excitation of $SH_0$ wave is examined by using the OSH-PT as the actuator and a $d_{36}$ type PMN-PT crystal as the sensor. Fig.7 shows the wave signals excited by the OSH-PT and received by the PMN-PT crystal at 180kHz along the 0° direction.

There is only one wave package in the received signal. By continuous wavelet transform (CWT) of the drive and received signals, the propagation time of the wave mode is determined to be 115.5 μs and the group velocity is calculated to be 3116 m·s$^{-1}$, which is very close to the theoretical velocity of SH$_0$ wave in aluminum plate (3099 m·s$^{-1}$). Since the PMN-PT wafer can receive both SH$_0$ wave and Lamb waves, and there is no other wave mode in the received signal in Fig.7(a), we can conclude that the two-half-ring OSH-PT can excite single mode SH$_0$ wave at 180kHz.

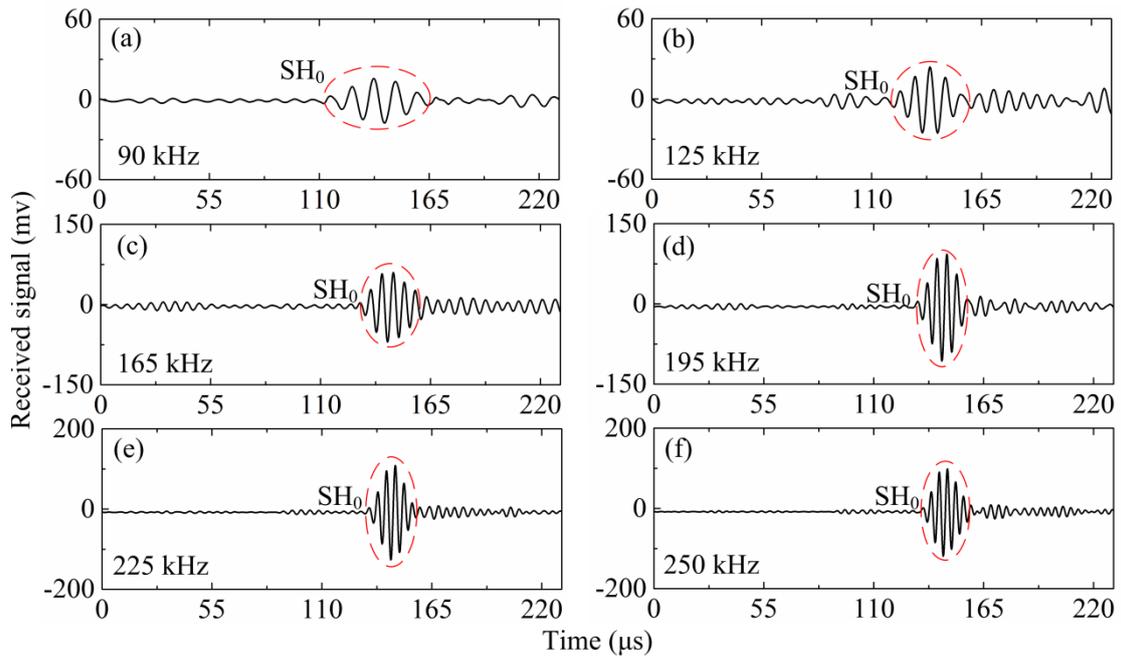

Fig.8. Wave signals excited by the two-half-ring OSH-PT at different frequencies with drive voltage of 120V and received by a d$_{36}$ type PMN-PT wafer.

The wave signals excited by the two-half-ring OSH-PT at different frequencies and received by a d$_{36}$ type PMN-PT wafer were plotted in Fig.8. It can be seen that the OSH-PT can excite single mode SH$_0$ wave in a wide frequency range from 90kHz to 250kHz. The SNR of the excited SH$_0$ wave increases steadily with the increasing frequency from 90kHz to 225kHz. Since the receiving properties of the PMN-PT wafer is also frequency dependent[22], the results in Fig.8 cannot accurately represent the frequency-dependent excitation performance of the two-half-ring OSH-PT, which can be realized by using an expensive 3-D laser vibrometer[32].

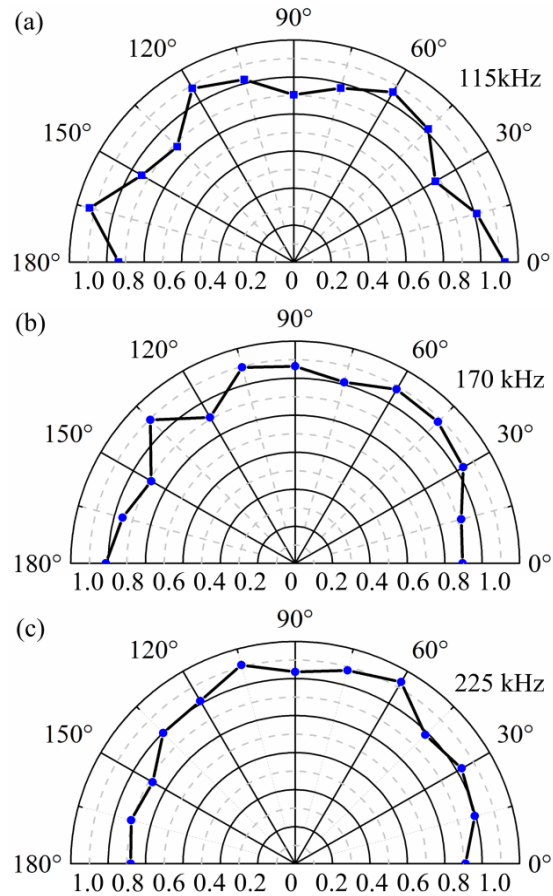

Fig.9. The omni-directivity of the two-half-ring OSH-PT in generation of $SH_0$ wave at (a) 115 kHz, (b) 170 kHz and (c) 225 kHz. Wave signals were excited by the OSH-PT and received by $d_{36}$ type PMN-PT single crystal wafers. The amplitude is normalized by average amplitude at their respective frequencies.

Fig.9 shows the omni-directivity of the two-half-ring OSH-PT in generation of $SH_0$ wave at different frequencies where the amplitude is normalized with regard to the average amplitude at the respective frequency. It can be seen that at all frequencies, the omni-directivity of the two-half-ring is quite good with the maximum deviation error of ~14%. Since the FEM simulation results had indicated that the theoretical omni-directivity of this OSH-PT is perfect, the deviation errors should come from the variations in properties of different PMN-PT wafers, bonding conditions of the actuator and sensors, etc. Also, the errors caused by sensors can be reduced or removed by measuring the full wave-field using a 3-D laser vibrometer.

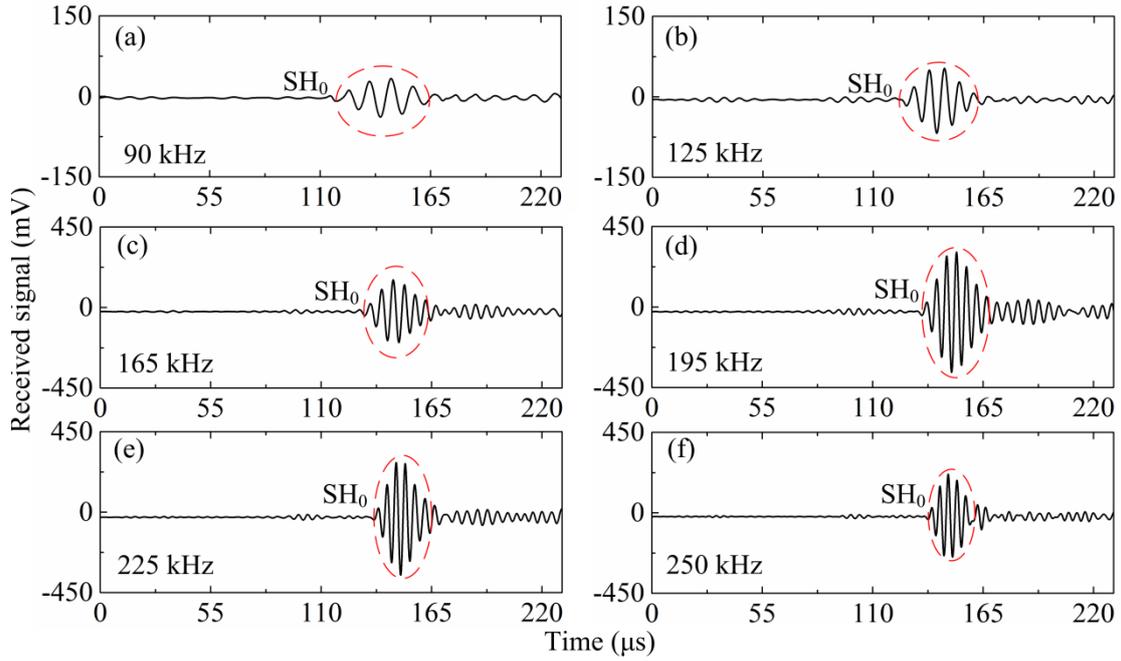

Fig.10. Wave signals excited by a $d_{36}$ type PMN-PT wafer at different frequencies with drive voltage of 50V and received by the two-half-ring OSH-PT.

Fig.10 shows the wave signals excited by a $d_{36}$ type PMN-PT wafer at different frequencies with the drive voltage of 50V and received by the two-half-ring OSH-PT. Due to the high sensitivity of the PMN-PT wafer in excitation, the received signals at all frequencies by the OSH-PT in figure 10 are obviously larger than that in Fig. 8, although the drive voltage in the former (50V) is lower than that in the latter (120V). Since the PMN-PT wafer can excite both $SH_0$ wave and Lamb waves[22], the single wave package in each of the subfigure indicates that the OSH-PT can filter some of the Lamb waves and receive the $SH_0$ wave in a wide range of frequency from 90kHz to 250kHz. Also, since the excitation properties of the PMN-PT wafer is also frequency dependent [22], the results in Fig.10 cannot represent the accurate reception properties of the OSH-PT. However, this problem cannot be solved at present since a frequency independent $SH_0$ wave generator does not exist at all.

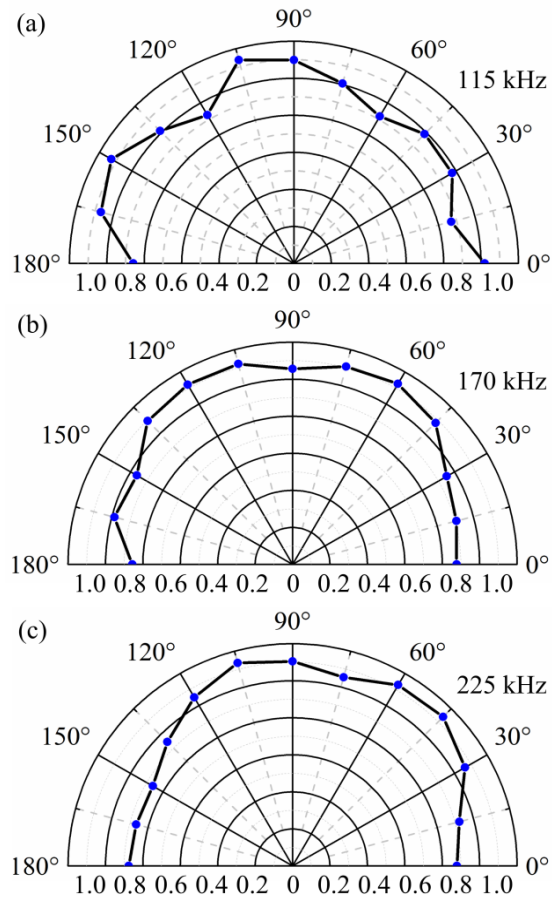

Fig.11. The omni-directivity of the two-half-ring OSH-PT in reception of $SH_0$ wave at (a) 115 kHz, (b) 170 kHz and (c) 225 kHz. Wave signals were excited by $d_{36}$ type PMN-PT wafers and received by the OSH-PT. The amplitude is normalized by average amplitude at their respective frequencies.

Fig.11 shows the omni-directivity of the two-half-ring OSH-PT in reception of $SH_0$ wave at different frequencies where the amplitude is also normalized with regard to the average amplitude at the respective frequency. It can be seen that the measured omni-directivity of the OSH-PT in reception is also quite good with the maximum deviations of ~14%. Also, the deviations should not be caused by the OSH-PT itself but from the PMN-PT actuators and bonding properties.

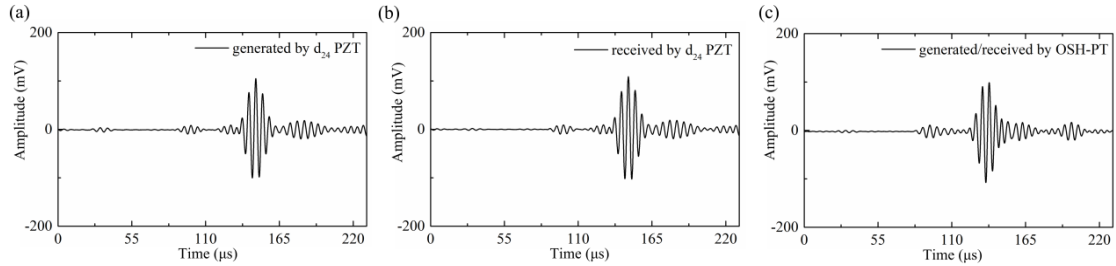

Fig.12. Received SH$_0$ wave signals from different actuator/sensor pairs at 180kHz with the drive voltage of 120V: (a) d$_{24}$ PZT/OSH-PT; (b)OSH-PT/d$_{24}$ PZT; (c)OSH-PT/OSH-PT

Finally, to examine the performances of the two-half-ring OSH-PT in practical testing, the wave signals obtained from three actuator/sensor pairs including d$_{24}$ PZT/OSH-PT, OSH-PT/d$_{24}$ PZT and OSH-PT/OSH-PT were measured at different frequencies. The former two actuator/sensor pairs can be used in tomography applications[37]. and the OSH-PT/OSH-PT pair is applicable to all the monitoring cases. In the testing, the in-plane poled d$_{24}$ PZT wafer is 8mm × 8mm × 1mm in dimension and electric field is applied in another in-plane direction (8mm). For all the testing, the drive voltage is fixed at 120V and the results at 180kHz were shown in Fig.12. It can be seen that in all the three cases, only one wave package (SH$_0$) appears in the received signal, indicating that all these three pairs of actuator/sensor can be used in practical testing. Meanwhile, the received amplitudes of SH$_0$ waves in these three cases were almost the same (~230mV), indicating the similar excitation/receiving sensitivity of the d$_{24}$ PZT wafer and the two-half-ring OSH-PT.

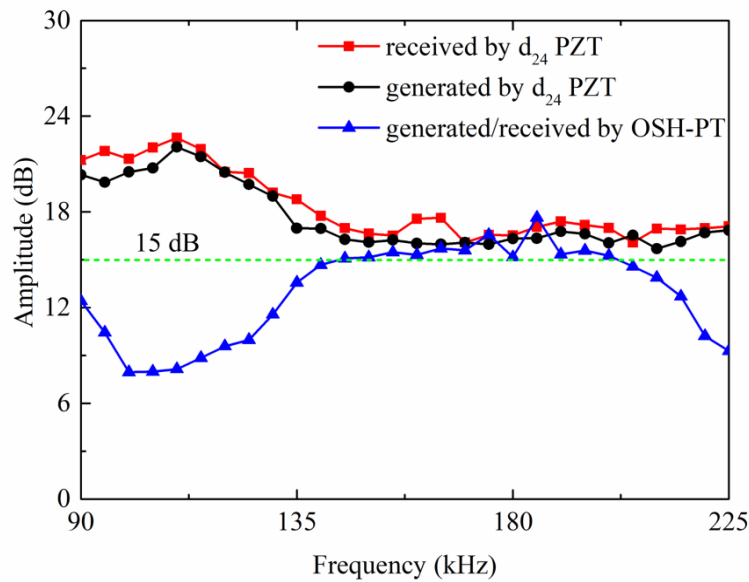

Fig.13. The frequency dependent SNR of the received wave signals for different actuator/sensor pairs. The drive voltage is 120V.

The SNR of the received wave signals for these three actuator/sensor pairs were measured from 90 kHz to 225 kHz and the results were shown in Fig.13. It can be seen that the SNRs for the actuator/sensor pairs of $d_{24}$ PZT/OSH-PT and OSH-PT/$d_{24}$ PZT are almost the same for all the testing frequencies, and they are overall higher than that for the actuator/sensor pair of OSH-PT/OSH-PT where one OSH-PT acts as actuator and another OSH-PT serves as sensor. Note that from 145kHz to 200kHz, the SNRs for these three actuator/sensor pairs are all above 15dB, which can be acceptable in most applications[38].

When comparing the performances of the two-half-ring OSH-PT fabricated in this work with the twelve-element OSH-PT reported previously[34], it is found that the SNR in the former (~15dB) is considerably lower than that of ~25dB in the latter, see Fig.4 in Ref.[35]. Therefore, the two-half-ring OSH-PT simplified the transducer structure at the sacrifice of reduced SNR. In practical applications, trade off should be made between the performance and fabrication/assembling complexity of the OSH-PT.

## 5. Conclusions

In summary, we demonstrated via FEM simulations that OSH-PTs based on thickness poled $d_{15}$ PZT rings with different number of elements can all show uniform omni-directivity in $SH_0$ wave excitation and reception. Overall, the simulated SNR in $SH_0$ wave excitation increases with the increasing number of elements, except for the six-element case. Due to its simplest structure, the two-half-ring OSH-PT was fabricated and its performances were systematically examined. Results show that its omni-directivity is fairly good with the maximum deviation of 14%. The SNR of the two-half-ring OSH-PT in self-excitation/self-reception can reach above 15dB from 145kHz to 200kHz, which is acceptable in most applications.

Although the FEM simulations had indicated that the SNR of the two-half-ring OSH-PT in excitation can be fairly enhanced by reducing the diameter of the ring, this has not been experimentally validated in this work due to the specimen limitations. The size effect of the two-half-ring OSH-PT in $SH_0$ wave excitation/reception will be systematically investigated in near future and an optimized dimension is expected to be obtained. Anyway, the two-half-ring OSH-PT presented in this work may greatly promote the applications of $SH_0$ wave based SHM due to its simple structure, easy fabrication/assembling and acceptable performances.

**Acknowledgements**

This work is supported by the National Natural Science Foundation of China under Grant Nos.11672003 and 11521202.

**References**

[1]     Halmshaw R 1991 *Non-destructive testing* (Arnold)

[2]     Balageas D, Fritzen C-P and Güemes A 2010 *Structural health monitoring* vol 90 (John Wiley & Sons)

[3]     Lynch J P and Loh K J 2006 A summary review of wireless sensors and sensor networks for structural health monitoring *Shock Vib. Dig.* **38** 91–130

[4]     Raghavan A and Cesnik C E 2007 Review of guided-wave structural health monitoring *Shock Vib. Dig.* **39** 91–116

[5]     Su Z, Ye L and Lu Y 2006 Guided Lamb waves for identification of damage in composite structures: A review *J. Sound Vib.* **295** 753–80

[6]     Staszewski W J 2004 Structural health monitoring using guided ultrasonic waves *Advances in smart technologies in structural engineering* (Springer) pp 117–62

[7]     Ihn J-B and Chang F-K 2008 Pitch-catch active sensing methods in structural health monitoring for aircraft structures *Struct. Health Monit.* **7** 5–19

[8]     Giurgiutiu V 2007 *Structural health monitoring: with piezoelectric wafer active sensors* (Academic Press)

[9]     Giurgiutiu V 2005 Tuned Lamb wave excitation and detection with piezoelectric wafer active sensors for structural health monitoring *J. Intell. Mater. Syst. Struct.* **16** 291–305

[10]    Yu L and Giurgiutiu V 2008 In situ 2-D piezoelectric wafer active sensors arrays for guided wave damage detection *Ultrasonics* **48** 117–34

[11]    Michaels J E 2008 Detection, localization and characterization of damage in plates with an in situ array of spatially distributed ultrasonic sensors *Smart Mater. Struct.* **17** 035035

[12]    Vasile C and Thompson R 1979 Excitation of horizontally polarized shear elastic waves by electromagnetic transducers with periodic permanent magnets *J. Appl. Phys.* **50** 2583–8

[13]    Thompson R B 1979 Generation of horizontally polarized shear waves in ferromagnetic materials using magnetostrictively coupled meander-coil electromagnetic transducers *Appl. Phys. Lett.* **34** 175–7

[14] Kwun H, Kim S and Crane J F 2002 Method and apparatus generating and detecting torsional wave inspection of pipes or tubes

[15] Kim I K and Kim Y Y 2006 Wireless frequency-tuned generation and measurement of torsional waves using magnetostrictive nickel gratings in cylinders *Sens. Actuators Phys.* **126** 73–7

[16] Rose J L 2014 *Ultrasonic guided waves in solid media* (Cambridge University Press)

[17] Alleyne D, Pavlakovic B, Lowe M and Cawley P 2004 Rapid, long range inspection of chemical plant pipework using guided waves Key Engineering Materials vol 270 (Trans Tech Publ) pp 434–41

[18] Cawley P, Lowe M, Alleyne D, Pavlakovic B and Wilcox P 2003 Practical long range guided wave inspection-applications to pipes and rail *Mater. Eval.* **61** 66–74

[19] Liu Z, He C, Wu B, Wang X and Yang S 2006 Circumferential and longitudinal defect detection using T (0, 1) mode excited by thickness shear mode piezoelectric elements *Ultrasonics* **44** e1135–8

[20] Wilcox P, Lowe M and Cawley P 2000 Lamb and SH wave transducer arrays for the inspection of large areas of thick plates AIP Conference Proceedings vol 509 (AIP) pp 1049–56

[21] Kamal A and Giurgiutiu V 2014 Shear horizontal wave excitation and reception with shear horizontal piezoelectric wafer active sensor (SH-PWAS) *Smart Mater. Struct.* **23** 085019

[22] Zhou W, Li H and Yuan F-G 2015 Fundamental understanding of wave generation and reception using d 36 type piezoelectric transducers *Ultrasonics* **57** 135–43

[23] Miao H, Dong S and Li F 2016 Excitation of fundamental shear horizontal wave by using face-shear (d36) piezoelectric ceramics *J. Appl. Phys.* **119** 174101

[24] Li F and Miao H 2016 Development of an apparent face-shear mode (d36) piezoelectric transducer for excitation and reception of shear horizontal waves via two-dimensional antiparallel poling *J. Appl. Phys.* **120** 144101

[25] Miao H, Huan Q and Li F 2016 Excitation and reception of pure shear horizontal waves by using face-shear d24 mode piezoelectric wafers *Smart Mater. Struct.* **25** 11LT01

[26] Miao H, Huan Q, Wang Q and Li F 2017 Excitation and reception of single torsional wave T (0, 1) mode in pipes using face-shear d24 piezoelectric ring array *Smart Mater. Struct.* **26** 025021

[27] Seung H M, Kim H W and Kim Y Y 2013 Development of an omni-directional shear-horizontal wave magnetostrictive patch transducer for plates *Ultrasonics* **53** 1304–8

[28] Seung H M, Park C I and Kim Y Y 2016 An omnidirectional shear-horizontal guided wave EMAT for a metallic plate *Ultrasonics* **69** 58–66

[29] Liu Z, Zhong X, Xie M, Liu X, He C and Wu B 2017 Damage imaging in composite plate by using double-turn coil omnidirectional shear-horizontal wave magnetostrictive patch transducer array *Adv. Compos. Mater.* **26** 67–78

[30] Borigo C J, Owens S E and Rose J L 2016 Piezoelectric shear rings for omnidirectional shear horizontal guided wave excitation and sensing

[31] Kim J O and Kwon O S 2003 Vibration characteristics of piezoelectric torsional transducers *J. Sound Vib.* **264** 453–73

[32] Belanger P and Boivin G 2016 Development of a low frequency omnidirectional piezoelectric shear horizontal wave transducer *Smart Mater. Struct.* **25** 045024

[33] Miao H, Huan Q, Wang Q and Li F 2017 A new omnidirectional shear horizontal wave transducer using face-shear (d 24) piezoelectric ring array *Ultrasonics* **74** 167–73

[34] Huan Q, Miao H and Li F 2017 A uniform-sensitivity omnidirectional shear-horizontal (SH) wave transducer based on a thickness poled, thickness-shear (d15) piezoelectric ring *Smart Mater. Struct.*

[35] Huan Q, Miao H and Li F 2018 A variable-frequency structural health monitoring system based on omnidirectional shear horizontal wave piezoelectric transducers *Smart Mater. Struct.* **27** 025008

[36] Miao H and Li F 2015 Realization of face-shear piezoelectric coefficient d36 in PZT ceramics via ferroelastic domain engineering *Appl. Phys. Lett.* **107** 122902

[37] Malyarenko E V and Hinders M K 2000 Fan beam and double crosshole Lamb wave tomography for mapping flaws in aging aircraft structures *J. Acoust. Soc. Am.* **108** 1631–9

[38] Alleyne D and Cawley P 1996 The excitation of Lamb waves in pipes using dry-coupled piezoelectric transducers *J. Nondestruct. Eval.* **15** 11–20